\documentclass[10pt] {article} %

\usepackage[T1]{fontenc} %
\usepackage[normalem]{ulem} %

\usepackage[usenames,dvipsnames,svgnames,x11names]{xcolor}

\usepackage{cite}

\usepackage{balance}

\usepackage{listings}

\lstdefinelanguage{XML}
{
basicstyle=\ttfamily\footnotesize,
  morestring=[b]",
  moredelim=[s][\bfseries\color{Maroon}]{<}{\ },
  moredelim=[s][\bfseries\color{Maroon}]{</}{>},
  moredelim=[l][\bfseries\color{Maroon}]{/>},
  moredelim=[l][\bfseries\color{Maroon}]{>},
  morecomment=[s]{<?}{?>},
  morecomment=[s]{<!--}{-->},
  commentstyle=\color{gray},
  stringstyle=\color{blue},
  identifierstyle=\color{red}
}

\usepackage{moreverb}

\usepackage[nounderscore]{syntax}

\usepackage[pdftex]{graphicx}
\graphicspath{{./figures/}}
\DeclareGraphicsExtensions{.pdf}

\usepackage[cmex10]{amsmath}
\usepackage{amssymb}
\usepackage{mathtools}
\usepackage{amsthm}
\usepackage{amsfonts}
\usepackage{gensymb}

\usepackage{subfig} %

\usepackage{algorithmicx}
\usepackage{algpseudocode}
\usepackage[ruled]{algorithm}
\definecolor{light-gray}{gray}{0.75}
\algrenewcommand{\algorithmiccomment}[1]{\hskip3em{{\footnotesize \textcolor{light-gray}{$\blacktriangleright$}}} #1}

\usepackage{multirow} %
\usepackage{rotating} %
\usepackage{booktabs} %
\usepackage{colortbl} %
\usepackage{tablefootnote} %

\usepackage{array}
\newcolumntype{L}[1]{>{\raggedright\let\newline\\\arraybackslash\hspace{0pt}}m{#1}}
\newcolumntype{C}[1]{>{\centering\let\newline\\\arraybackslash\hspace{0pt}}m{#1}}
\newcolumntype{R}[1]{>{\raggedleft\let\newline\\\arraybackslash\hspace{0pt}}m{#1}}

\usepackage[pdftex,colorlinks=true,urlcolor=blue,citecolor=blue]{hyperref}

\usepackage{xspace}

\usepackage{enumitem}

\usepackage{blindtext}

\newcommand{\opes}{OpES\xspace}
    
\begin{document}

\title{Optimizing Federated Learning using Remote Embeddings for Graph Neural Networks
\thanks{~Preprint of paper in the proceedings of the 30th International European Conference on Parallel and Distributed Computing (Euro-Par): Pranjal Naman and Yogesh Simmhan, “Optimizing Federated Learning using Remote Embeddings for Graph Neural Networks,” in \textit{International European Conference on Parallel and Distributed Computing (Euro-Par)}, 2024}
}

\author{Pranjal Naman and Yogesh Simmhan\\
Department of Computational and Data Sciences (CDS),\\
Indian Institute of Science (IISc),\\
Bangalore 560012 India\\
Email: \{pranjalnaman, simmhan\}@iisc.ac.in
}

\date{}
\maketitle

\begin{abstract}
Graph Neural Networks~(GNNs) have experienced rapid advancements in recent years due to their ability to learn meaningful representations from graph data structures. 
Federated Learning~(FL) has emerged as a viable machine learning approach for training a shared model on decentralized data, addressing privacy concerns while leveraging parallelism. 
Existing methods that address the unique requirements of federated GNN training using remote embeddings to enhance convergence accuracy
are limited by their diminished performance due to large communication costs with a shared embedding server. In this paper, we present \opes, an optimized federated GNN training framework that uses remote neighbourhood pruning, and overlaps pushing of embeddings to the server with local training to reduce the network costs and training time. The modest drop in per-round accuracy due to pre-emptive push of embeddings is out-stripped by the reduction in per-round training time for large and dense graphs like Reddit and Products, converging up to $\approx 2\times$ faster than the state-of-the-art technique using an embedding server and giving up to $20\%$ better accuracy than vanilla federated GNN learning.
\end{abstract}

\section{Introduction}
Graph Neural Networks~(GNNs) have emerged as a powerful tool for learning representations on non-Euclidean data, effectively capturing dependencies and relational information inherent in them~\cite{kipf2016semisupervised}. GNNs leverage the graph topology as well as the node and edge features to learn low-dimensional embeddings, enabling them to perform tasks such as node classification, edge prediction and graph classification~\cite{kipf2016semisupervised, morris2019weisfeiler}. 

When training a \textit{k}-layered GNN architecture, the \textit{forward pass} aggregates the \textit{k}-hop neighbours and their features/embeddings for each \textit{labelled target vertex} in the graph, while the \textit{backward pass} propagates the loss relative to the predicted label among the layers to update the weights and embeddings in the neural network. This is done iteratively, with each \textit{mini-batch} training on a subset of target vertices and a training \textit{epoch} ensuring coverage over all target vertices. In doing so, the embeddings generated capture both the graph topology as well as features on the neighbouring vertices or/and edges. 

\paragraph{Limitations of Centralized Training.}
Real-world graph datasets can be substantially large, e.g., buyers and products in an eCommerce site, interactions between user accounts in a fintech transaction graph, etc., making it computationally costly and memory-intensive to perform GNN training on a single~(even accelerated) server. Additionally, strict data privacy regulations, such as GDPR
, can prohibit data owners from sharing private information or aggregating their local interaction graphs centrally to a cloud server outside certain geographies or their institutions. However, applications may wish to train GNN models over graphs that span such distributed locations. For instance, a global eCommerce seller may wish to use product purchase trends by users in one country to recommend the product to users in other countries, or banks may want to detect frauds in transactions, even when they span two banking institutions.

\paragraph{Federated learning of GNNs.}
Federated Learning~(FL)~\cite{mcmahan2017fedavg} has recently shown promise for training Deep Neural Network~(DNN) models across data present at multiple clients~(devices, workstations) without the need to move them to a central server~(cloud). FL trains DNN models locally over data present in each client, shares these local models with a central server, which aggregates them into a single global model for one \textit{round} of training, and sends the global model back to the clients for another round of training, until the global model converges. Thus, the global model incorporates training features from data present across the clients while not sharing the original data with the server. The success of FL in fields like computer vision and natural language processing has inspired its application for GNNs to avoid or limit the sharing of graph data from a client to a central server or with other clients~\cite{wang2022fedscope, wu2023embc, yao2023fedgcn}.

\begin{figure}[!t]
    \subfloat[Design of Vanilla and EmbC (with embedding database) Federated GNN training.]{\label{subfig:arch}
      \includegraphics[width=0.35\textwidth]{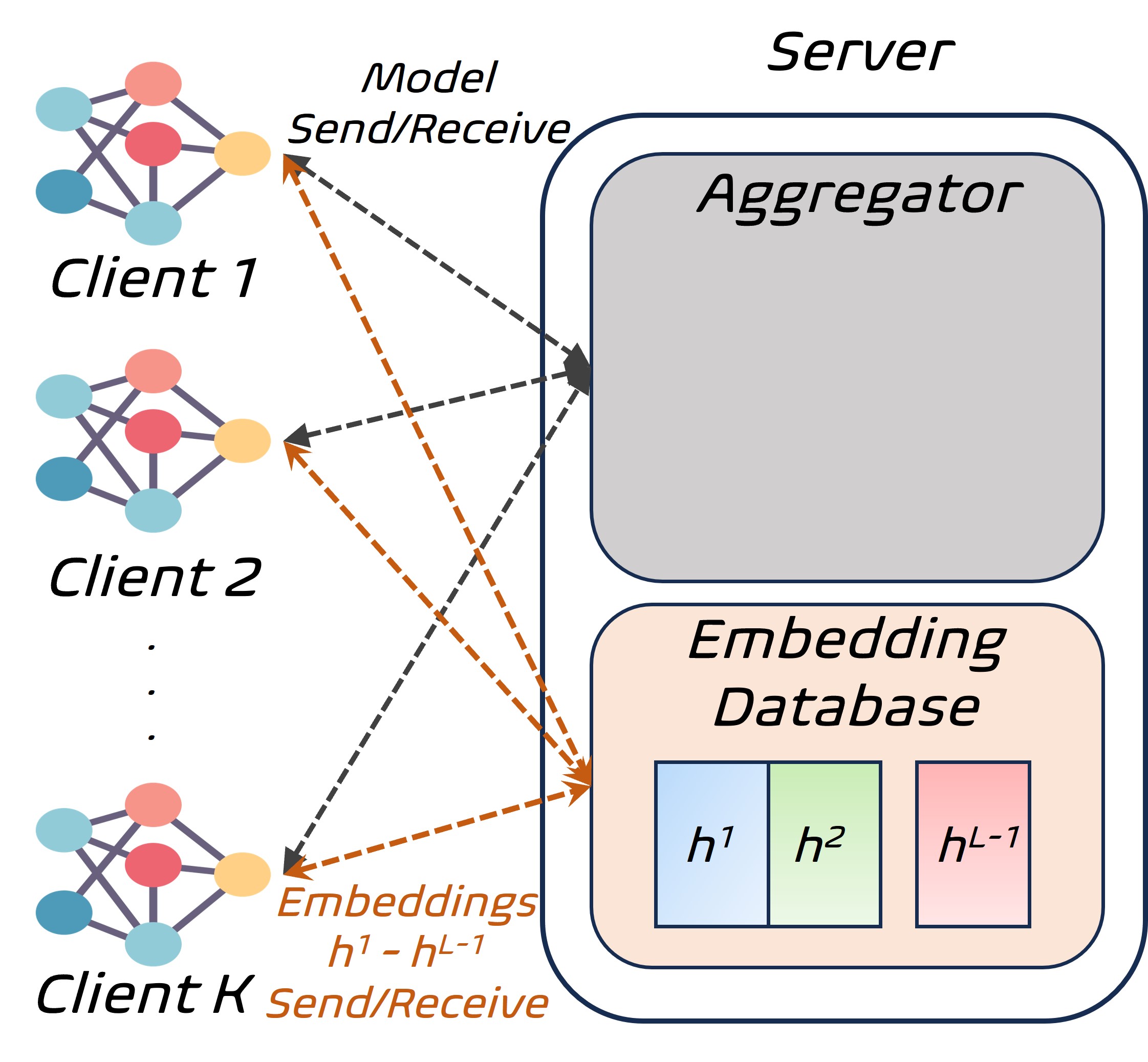}
    }~
    \subfloat[\% of remote vertices in subgraph partitions (left, bar) and \# of embeddings stored (right, marker).
    \label{subfig-2:inside-outside}]{%
      \includegraphics[width=0.26\textwidth]{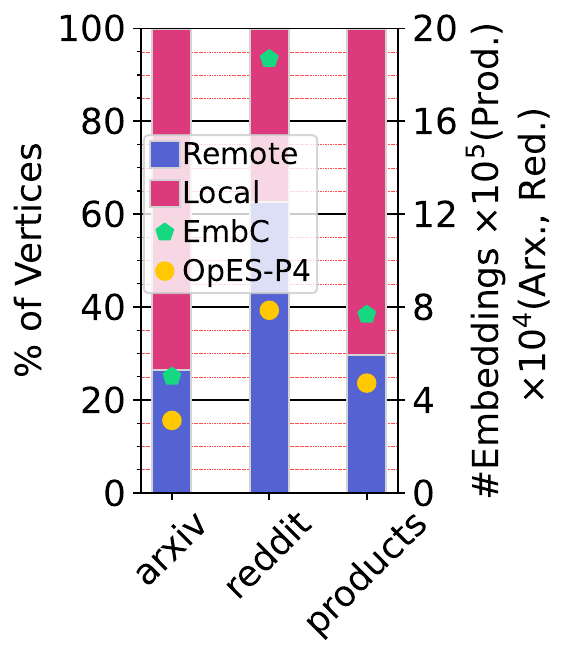}
    }~
    \subfloat[Time-to-accuracy~(TTA) for our \opes over EmbC and Vanilla federated GNN training to reach the nominal peak accuracy achieved by VFL.
    \label{subfig-2:motivation3}]{%
      \includegraphics[width=0.34\textwidth]{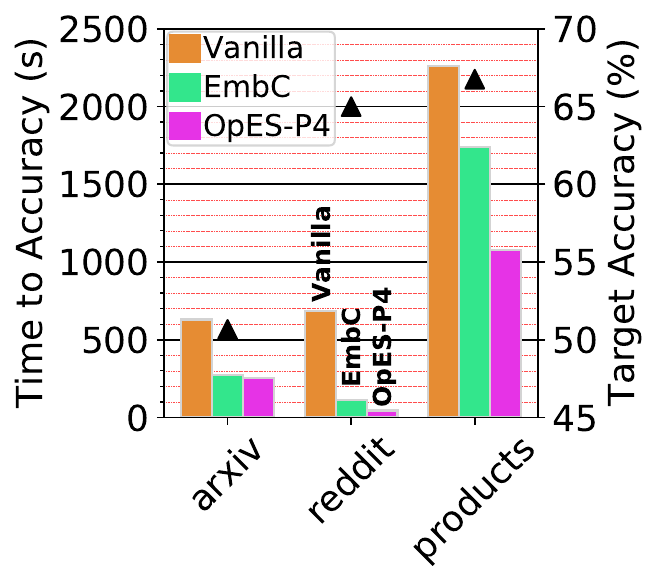}
    }
    \caption{Baseline design for federated learning of GNNs and need for optimizations.}
    \label{fig:motivation}
  \end{figure}

FL on GNNs is divided into three categories, based on the graph and linkages between them maintained on the clients: (1) \textit{Ego-network level}, where each client holds the \textit{n}-hop neighbourhood for one or more labelled vertices, e.g., interactions between the user and other entities on a smartphone client; (2) \textit{Subgraph level}, where each client holds a partitioned subgraph of a larger graph, e.g., transactions pertinent to one bank~(client) for a transaction graph that spans all national banks; and (3) \textit{Graph level}, where each client stores one or more complete, independent graphs, e.g., protein molecule graphs held by multiple pharmaceutical companies.
In this paper, we focus on GNN training over \textit{partitioned subgraphs} that span different clients, as found in transaction networks across institutions or enterprise data across regulatory boundaries. Here, a GNN model is trained locally on the subgraph present within a client and the local models from different clients are aggregated on a central server. We refer to this as \textit{vanilla} federated GNN training (VFL, Fig.~\ref{subfig:arch}). 

\paragraph{Challenges and Gaps of Federated GNN Training.}
Federated learning of a GNN model over linked data presents unique challenges compared to traditional FL over independent training samples. A key limitation is the dependency between data present across clients, specifically when subgraphs situated on different clients share \textit{cross-client edges} that link their vertices, e.g., a transaction edge between two user account vertices in two different banks. Since GNN models train over the neighbourhood information for a labelled vertex, these cross-client edges and their downstream vertices form part of the training neighbourhood. Omitting them may result in a significant degradation of the model's performance, but including their raw features can compromise privacy. For instance, in our experiments, the Reddit graph partitioned onto 4 clients achieves only 60\% training accuracy when cross-client edges are omitted~(VFL) but achieves 80\% accuracy with partial embedding information shared, as we discuss later.

Prior works have attempted to address this limitation by using a subset of cross-client neighbourhood information. \textit{FedSAGE}~\cite{zhang2021fedsage} proposes a generative model to generate cross-client neighbours and their features, trained simultaneously with the GNN model but at the cost of increased training time. \textit{FedGCN}~\cite{yao2023fedgcn} proposes a pre-training round to get the aggregated features of all the cross-client neighbours, but this can leak private data since the aggregated features for $1$-hop and $2$-hop are communicated directly. \textit{EmbC-FGNN}~\cite{wu2023embc} introduces a promising approach of sharing just the embeddings for the vertices incident on the cross-client edges through an embedding server. This restricts data sharing to just the anonymized embeddings for remote vertices rather than their raw features while allowing neighbourhood training to benefit from them. Their clients push the updated cross-client embeddings to an \textit{embedding server} after each round of training and pull the relevant ones at the start of the next round. This performs better than other federated graph learning strategies that they compare against. 

However, a notable drawback of the EmbC method is the communication cost for sharing potentially $100,000s$ of embeddings in each round between the clients and the embedding server, and a large in-memory footprint of the embeddings maintained at the server. This is demonstrated in Fig.~\ref{subfig-2:inside-outside}, which shows the fraction of local and remote vertices~(i.e., unique vertices on a different client connected through a cross-client edge from this client) for three common GNN graphs we evaluate, Arxiv, Reddit and Products, having $0.17$--$2.4M$ vertices and $1.16$--$123.7M$ edges and each split into $4$ partitions. $15$--$30\%$ of their vertices are connected to a remote vertex, indicating their influence on the training neighbourhood, and this translates to over $\approx800k$ embedding vectors maintained in the embedding server and transferred between the server and clients in each training round.

\paragraph{Approach and Contributions.}
In this paper, we propose the \textbf{Optimized Embedding Server (OpES)}, an enhanced approach that reduces the communication, computation and memory costs for the shared embeddings method, which leads to faster training time per round, faster training time to convergence, minimum to no reduction in accuracy, and no additional sharing of data. Specifically, we make the following contributions: 
\begin{enumerate}
    \item We optimize the communication cost by \textit{overlapping} the transfer of embeddings from clients to the embedding server with the final epoch of training in a round, thus hiding the network costs while using slightly stale embeddings.
    \item We \textit{prune} the neighbourhood of \textit{remote} vertices that are used during training, reducing both the embeddings transferred and the in-memory footprint, and also reducing the compute cost for forward and backward passes. 
    \item These are validated by detailed experiments on three real-world graphs and comparison with vanilla FL and the state-of-the-art (SOTA) EmbC approaches. These lead to $38\%$ fewer embeddings maintained for the Products graph (Fig.~\ref{subfig-2:inside-outside}), dropping from $768k$ for EmbC to $473k$ for \opes, while we reduce the time to accuracy (Fig.~\ref{subfig-2:motivation3}) from $2260s$ for Vanilla and $1739s$ for EmbC to $1076s$ for \opes for the same graph.
\end{enumerate}

In the rest of the article, we discuss related works in \S~\ref{sec:related}, present our proposed optimization methods for \opes in \S~\ref{sec:methods}, offer detailed comparative experiments using different configurations in \S~\ref{sec:results}, and summarize our conclusions in \S~\ref{sec:conclude}.

\section{Related Works}
\label{sec:related}

Federated learning on graphs has gained traction recently~\cite{yao2023fedgcn, zhang2021fedsage}.
This entails training a global GNN model in a decentralized manner while restricting the graph data to each data owner~(client). While the concept is quite similar to traditional federated learning, federated learning over graphs introduces an added layer of complexity. In tasks such as semi-supervised node classification, the data local to the clients may exhibit inter-client dependencies, which must be addressed in a privacy-respecting manner to achieve improved model convergence. The exchange of limited information to account for these inter-client dependencies can lead to large communication overheads, and if not done carefully, risk privacy leakage. Several recent works have tried to address these issues. 

Many methods target subgraph-based federated graph learning, particularly addressing the challenge of missing details on cross-client edges on remote clients that affect training accuracy. Some tackle this by expanding their local subgraphs. \textit{FedSAGE}~\cite{zhang2021fedsage} performs this expansion by generating the neighbourhood and its features through a collaboratively learned generative model. Firstly, all clients train a neighbourhood generator model jointly by sharing gradients. The clients then use this neighbourhood generator model to expand their subgraphs and conduct a localized federated graph learning round. In contrast, \textit{FedGNN}~\cite{wu2021fedgnn} achieves expansion by exactly augmenting the relevant nodes from other subgraphs. In addition to solving the issue of handling cross-client edges, other works also look at challenges such as data heterogeneity and improved aggregation techniques for topology-aware model aggregation at the server. \textit{FedPUB}~\cite{baek2023fedpub} proposes a personalized subgraph federated learning to tackle distribution heterogeneity of data, particularly the fact that the subgraphs on data owners might belong to different communities in the global graph and a naive FL algorithm would lead to poor performance. They propose calculating the similarity between subgraphs and grouping similar subgraphs for improved performance. \textit{GCFL}~\cite{xie2021gcfl} is a federated graph classification framework that also tackles the non-IID nature of the participating subgraphs and the node features. GCFL observes that most real-world graphs share at least a few similar properties, such as largest component size and average shortest path lengths and calculates similarities among subgraphs to cluster similar clients for a better global model convergence. 
\vspace{-0.032cm}

\textit{EmbC}~\cite{wu2023embc} introduces a SOTA subgraph federated learning based using subgraph expansion to include the latest embeddings for the cross-client neighbours at the boundary between rounds. A trusted entity called the \textit{embedding server} informs the clients of the presence of cross-client neighbouring vertices. The clients expand their local subgraphs to include these remote vertices, and use them while training. Specifically (Fig.~\ref{subfig:arch}), 
at the start of a round, each client \textit{pulls} the embeddings from the embedding server for such remote vertices they have \textit{edges to} their local vertices, and also \textit{pushes} the updated embeddings to the embedding server for the local vertices which have \textit{edges to} remote vertices in other clients -- these updated embeddings will be pulled and used by those other clients in the next round. 

However, EmbC suffers from diminished performance due to numerous embeddings maintained in the server as the 
GNN model depth increases and/or the number of partitions increases, and hence, the cross-client edge cuts increase. This causes the memory footprint of the embedding server to grow~(since it maintains these in-memory for fast response), and also the communication cost between client and server to grow in each round. 
We address this limitation in \opes by expanding the local subgraph in a pruned fashion and overlapping the embeddings communication and model computation within a round, achieving faster model convergence without tangibly reducing the peak accuracy achieved.

\section{\opes Design and Optimizations}
\label{sec:methods}

In this section, we outline our training approach and the proposed optimizations to expedite the federated learning process.

\subsection{Architecture}

\begin{figure}[t]
    \centering
    \includegraphics[width=0.7\textwidth]{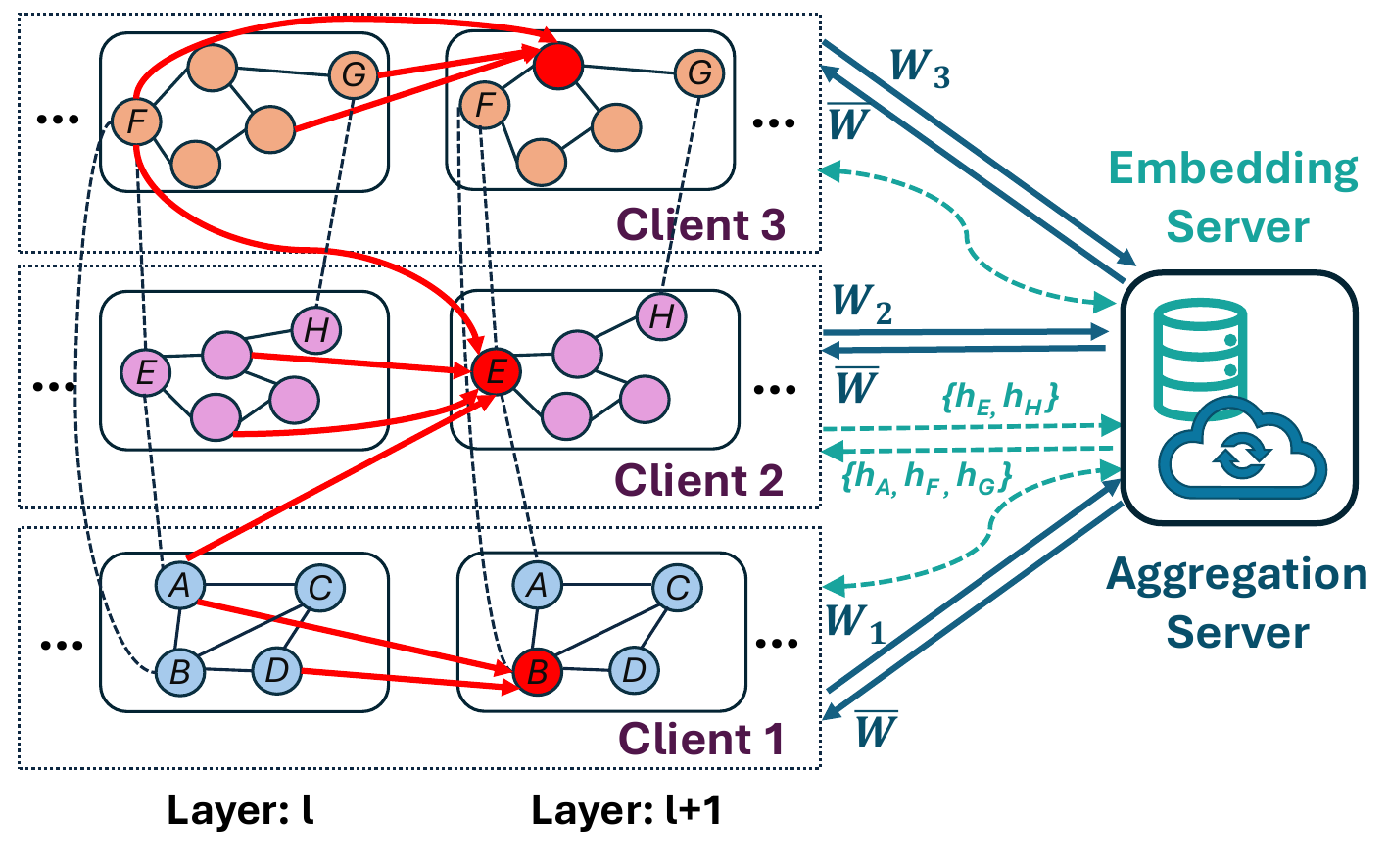}
    \caption{Federated GNN training flow using remote embeddings. Before training, each client pulls embeddings for its \textit{pull nodes}~(shown for Client 2). Training involves nodes in a mini-batch~(red nodes) aggregating neighbours' embeddings~(shown in red arrows) to generate the next layer's embeddings. After training, clients upload their local model weights to the aggregation server and the updated embeddings of their \textit{push nodes} to the embedding server. The aggregation server then aggregates these weights into the updated global model and distributes it back to all clients.}
    \label{fig:fgnn-trainl}
\end{figure}

The base architecture of \opes is similar to EmbC~\cite{wu2023embc} and consists of three key components: a \textit{central aggregation server} that orchestrates the federated learning rounds, performs client selection, and aggregates the models; a~(possibly co-located) \textit{embedding server} that is aware of the cross-client edges and receives/sends the embeddings for the remote vertices maintained on it from/to the relevant clients; and the \textit{clients} that perform the local training rounds, interacting with the aggregation and embedding servers. We describe these in detail.

The \textit{aggregation server} is a typical FL server that can perform any number of client selection or model aggregation strategies such as FedAvg~\cite{mcmahan2017fedavg}, TiFL~\cite{chai2020tifl}, among others. The \textit{clients} initially possess only the local subgraph partitions of the overall graph and can query the \textit{embedding server} for the presence of cross-client neighbours~(remote vertices). Once discovered, the clients expand their local subgraphs to include these remote neighbours present in other clients, tagging them with a flag indicating they are remote. The embedding server is an \textit{honest-but-curious} entity. So, 
only the structure of the cross-client 1-hop vertices in the graph~(i.e., just their vertex IDs) is shared with the clients; neither the embedding server nor the clients have access to the features of the remote vertices or the remote edges located on other clients. This standard privacy model is followed by several federated graph learning frameworks~\cite{yao2023fedgcn}.
The embedding server maintains an in-memory Key-Value~(KV) store that stores $L - 1$ embeddings~($h^1$ to $h^{L-1}$ for an $L$ layered GNN) for all vertices whose embeddings need to be shared to/from the relevant clients. The same remote vertex may be present at a 1-hop, 2-hop, etc. distances from a target labelled vertex that is part of a training mini-batch. So, each hop-distance translates to a different embedding for that vertex~(Fig.~\ref{fig:comp-graph:b}).

\subsection{Lifecycle of a Training Round}

Fig.~\ref{fig:fgnn-trainl} shows the standard training life cycle of a federated subgraph learning using remote embeddings.
We first establish standard terminology that will be used going forward. For any client $k$, local vertices that are neighbours of another client~(i.e., remote vertices for those clients) and are required for completing their training are referred to as the \textit{push nodes} of client $k$. Similarly, vertices whose embeddings are necessary to complete a training round are termed \textit{pull nodes} of client $k$. We assume that the GNN model being collaboratively trained has $L$ layers, and we run $\epsilon$ epochs in each round of training. 

\begin{figure}[t!]
\centering
      \subfloat[Subgraphs on different clients. Cross-client edges are shown as dashed edges.]{\label{fig:comp-graph:a}
  \quad\qquad\includegraphics[width=0.32\textwidth]{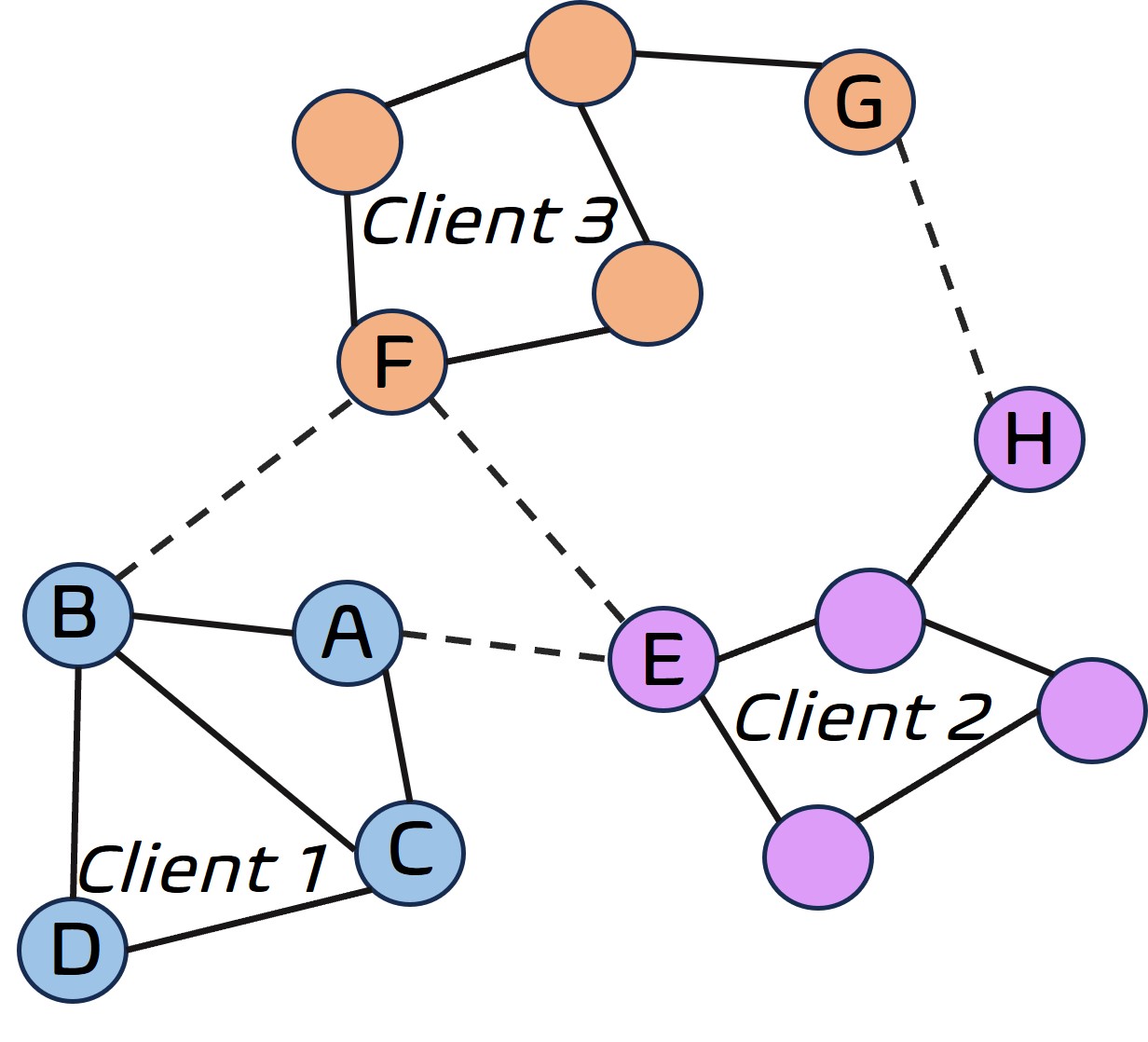}\qquad\quad
    }~~
      \subfloat[Computation graph generated for training vertex A on client 1]{\label{fig:comp-graph:b}
  \quad\qquad\includegraphics[width=0.3\textwidth]{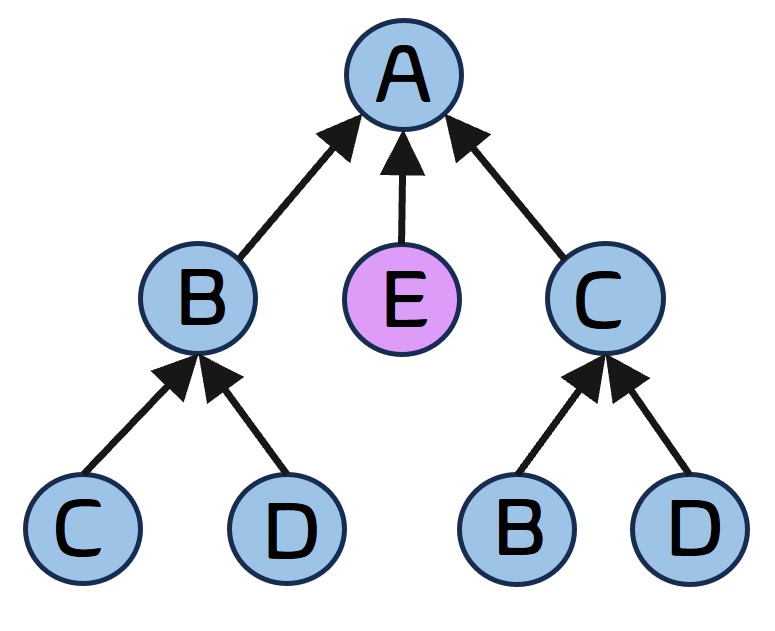}\qquad\quad
    }
\caption{Existence of cross-client edges and its impact on computation graph for training node \textit{A}~(for a 2-layered GNN)}
\label{fig:comp-graph}
\end{figure}

\paragraph{Pre-training.}
Before the iterative training rounds start, we run a pre-training round to initialize the embeddings for all \textit{push nodes} in each client. This occurs on the local subgraph~(before its expansion), calculating the $h^1$ to $h^{L-1}$ embeddings of the client's push nodes. Client 1 in Fig.~\ref{fig:comp-graph:a} initializes the $h^1$ embeddings for $\{ A, B \}$ since these are needed by other clients. These embeddings are then sent to the embedding server for use in subsequent rounds. The pre-training round takes place once per FL session.

\paragraph{Sampling \& Training.}

Each round commences with a \textit{pull phase}, where each client fetches embeddings from the embedding server for the \textit{pull nodes} to be utilized in the current round. As shown in Fig~\ref{fig:comp-graph:a}, Client 1 pulls the $h^1 - h^{L-1}$ embeddings for nodes $F$ and $E$. Node features~($h^0$ embeddings) are unavailable at the embedding server due to privacy reasons and cannot be pulled. In this case, the GNN has $L=2$ layers; therefore, only the $h^1$ embeddings are pulled. Similarly, at the start of their individual rounds, Clients 2 and 3 pull the $h^1$ embeddings for vertices $\{ A, F, G \}$ and $\{B, E, H \}$, respectively. The pulled embeddings are cached in-memory locally on the client for use in this round.

GNN training is typically done in \textit{minibatches} due to the high memory requirements of full-batch training. A subset of training nodes is selected in each minibatch, and the computation graph for these is built using a \textit{neighbourhood sampler}. The sampler samples a fixed number of neighbouring nodes at each hop of the training nodes to build the computation graph.
For the \textit{sampling phase}, we develop a custom sampler that ignores the cross-client edges at the penultimate layer. The custom sampler generates the computation graph, keeping in mind the following points. Firstly, only local nodes are sampled at the root level. This is the default behaviour. Between hops $1$ to $L-1$, any local or remote vertex can be sampled. However, once a remote node is sampled at hop $l$, the path does not grow any further~(e.g., node $E$ in Fig.~\ref{fig:comp-graph:b}). The custom sampler also ensures that no remote nodes appear at the $L^{th}$ hop since the $h^0$ embeddings~(i.e., the features) of the remote nodes are unavailable. The node embeddings do not reveal any information about the specific neighbours of the remote node and only represent an aggregated view. Hence, it is safe to share embeddings among clients. 

The \textit{forward pass} of local training commences after sampling. Here, each layer \textit{l} takes the $h^{l-1}$ embeddings of the nodes at the $(L - (l - 1))^{th}$ hop as input and generates the $h^{l}$ embeddings of the nodes at $(L - l)^{th}$ hop. Before this is done, the $h^{l-l}$ embeddings for the remote vertices at the $(L - (l - 1))^{th}$ hop are loaded from the local cache. This process repeats for all the layers of the GNN to generate the $h^L$ embeddings at the output layer. The loss is calculated at the end of the forward pass and is propagated backwards along the neural network in the \textit{backwards pass}. The tasks of \textit{sampling}, \textit{forward pass} and the \textit{backward pass} are repeated for all minibatches in an epoch of local training.

The \textit{push phase} begins after all the epochs of a round have completed. Here, the locally trained model is used to calculate the $h^1$--$h^{L-1}$ embeddings are calculated for the push nodes in the same manner \textit{forward pass} takes place. Once all the embeddings have been calculated for all the push nodes, these embeddings are pushed to the embedding server.

\subsection{Pruning Optimization}
Full-batch training of GNNs has high memory requirements and suffers from the neighbourhood explosion problem. There is an exponential growth in the number of vertices and edges encountered when traversing the neighbourhood of a vertex for every additional hop in a large and dense graph. To circumvent this, GNN training is generally done in mini-batches, which has been shown to reduce the memory requirements of training. Hamilton et al.~\cite{hamilton2017graphsage} proposed to sample a fixed-size neighbourhood at each hop to reduce the memory requirements while still achieving great results. 
With this consideration in mind, we limit the expansion of the local subgraph to not fully encompass all remote vertices. 

Instead, we propose a pruning technique to \textit{restrict the maximum number of remote vertices} in the expanded subgraph. Doing so can accelerate the convergence time for the local round while ensuring that the local GNN model does not completely disregard the cross-client edges. This reduction in time per round can arise due to multiple factors. Firstly, there is a sharp reduction in the embeddings of the remote vertices that need to be pulled from the server and cached locally in each training round, thereby decreasing the communication in the pull phase of the round. Additionally, the training time is shortened because the forward pass encounters fewer remote vertices, and hence has to access and populate fewer embeddings from the local cache. Lastly, pruning the expanded subgraphs also reduces the number of embeddings for local push vertices that must be sent to the embedding store at the push phase of the round. This pruning is performed at random, i.e., a random subset of neighbouring remote vertices are removed to stay within the \textit{retention limit}; for simplicity, this is done offline before loading the subgraph in our implementation.

We later demonstrate an ablation study on pruning using different numbers $i$ of \textit{remote vertices retained} after pruning~($P_i$). It shows that such pruning has only a marginal effect on the peak accuracy, and yet decreases the per-round time. As is intuitive, retaining zero vertices after pruning~($P_0$) reduces this approach to a vanilla federated GNN with no embeddings shared; having a retention limit that is unbounded ($P_\infty$) means no vertices are pruned similar to the baseline EmbC model.

\subsection{Overlapping Push with Compute Optimization}
Another optimization we propose further hides the time taken by the push phase of a round. A round typically consists of multiple epochs of local training so that the local model stabilizes its learning. We use the intuition that the embeddings for the push vertices that are computed are unlikely to change by much between the end of the last epoch $\epsilon$ and the penultimate epoch $(\epsilon - 1)$. This gives us the opportunity to
schedule a send of the embeddings for the push vertices' states at the end of epoch $(\epsilon - 1)$ while training of the final epoch is ongoing. The push is communication bound while the training is compute bound; hence, this overlapping of push and training phases allows complementary resources to be used and the push phase time to be hidden within the training phase time, albeit with a slightly stale version of the embeddings. 

Specifically, at the end of the epoch $(\epsilon - 1)$ epoch in a round, we launch a separate process that takes the trained local model until that epoch and performs a forward pass to generate the embeddings for layers $h^1$--$h^{L-1}$ of the push vertices, and sends these to the embedding server. The local model will continue to train for the last epoch $\epsilon$ in the round and will be sent to the aggregation server for global model aggregation. This optimization is relevant only when $\epsilon \geq 2$.

\subsection{Implementation}
We implement the vanilla federated GNN and the EmbC-FGNN framework, along with all our optimizations, using DGL~\cite{wang2019dgl} and PyTorch and the Flotilla~\cite{flotilla} federated learning framework. The embedding store is implemented as a Redis server that stores $\langle key, value \rangle$ pairs with the $key$ being the node ID, and the $value$ being the corresponding embedding. We implement efficient, pipelined \textit{batch get} and \textit{set} operations on the Redis datastore to pull and push embeddings. 

\section{Experiments \& Results}
\label{sec:results}

\begin{table}[t!]
\centering
\setlength{\tabcolsep}{2pt}
\def\arraystretch{0.9}
\footnotesize
\caption{Graph Datasets Used for Experiments}
\begin{tabular}{l||r|r|r|r|r|r}
\hline
\textbf{Graph} & $|V|$  & $|E|$   & Features & \# classes & Avg. Degree & Train Nodes \\ \hline\hline
\textbf{\textit{ogbn-arxiv   }}                & 169.3K   & 1.17M  & 128 & 40 & 13.7 & 90.94K \\
\textbf{\textit{reddit}}               & 233K   &  114.85M  & 602 & 41 & 492 & 153.43K \\
\textbf{\textit{ogbn-products }}          & 2.45M   & 123.72M & 100  & 47 & 50.5 &  196.62K\\ \hline
\end{tabular}
\label{tab:dataset-specs}
\end{table}

\subsection{Experimental Setup}
Our experimental setup consists of four GPU workstation \textit{clients} with an Nvidia RTX 4090~(24GB) GPU card, an AMD Ryzen 9 7900X with 12-Core CPU and $128$GB RAM each, and a GPU server with an Nvidia RTX A5000~(24GB) GPU card, an AMD EPYC 7532 $32$-Core CPU and $512$GB of RAM. The server hosts the aggregation and embedding store services, the latter using the Redis Key-Value store. 
The machines are connected using a gigabit Ethernet interface.

We use three common GNN datasets for our experiments: \textit{ogbn-arxiv}, \textit{reddit}, and \textit{ogbn-products}, which represent a citation network from the arXiv paper repository, Reddit social network posts, and products from an eCommerce website~(Table~\ref{tab:dataset-specs}). We use the METIS~\cite{karypis1997metis} partitioner to divide the graphs into $4$ subgraphs with vertex balancing and minimized edge cuts, and each subgraph is assigned to a client. Our experiments use a $3$-layer GraphConv~\cite{kipf2016semisupervised} GNN model with a hidden embedding size of $32$. The epochs per round are set to $\epsilon=3$, and the learning rate to $0.001$. The batch sizes selected for Arxiv, Reddit, and Products are $64$, $1024$, and $2048$, respectively. The accuracies are measured on a global test dataset held by the aggregation server.

\subsection{Benefits of Push Overlap Optimization}

\begin{figure}[t!]
    \centering
    \includegraphics[width=\textwidth]{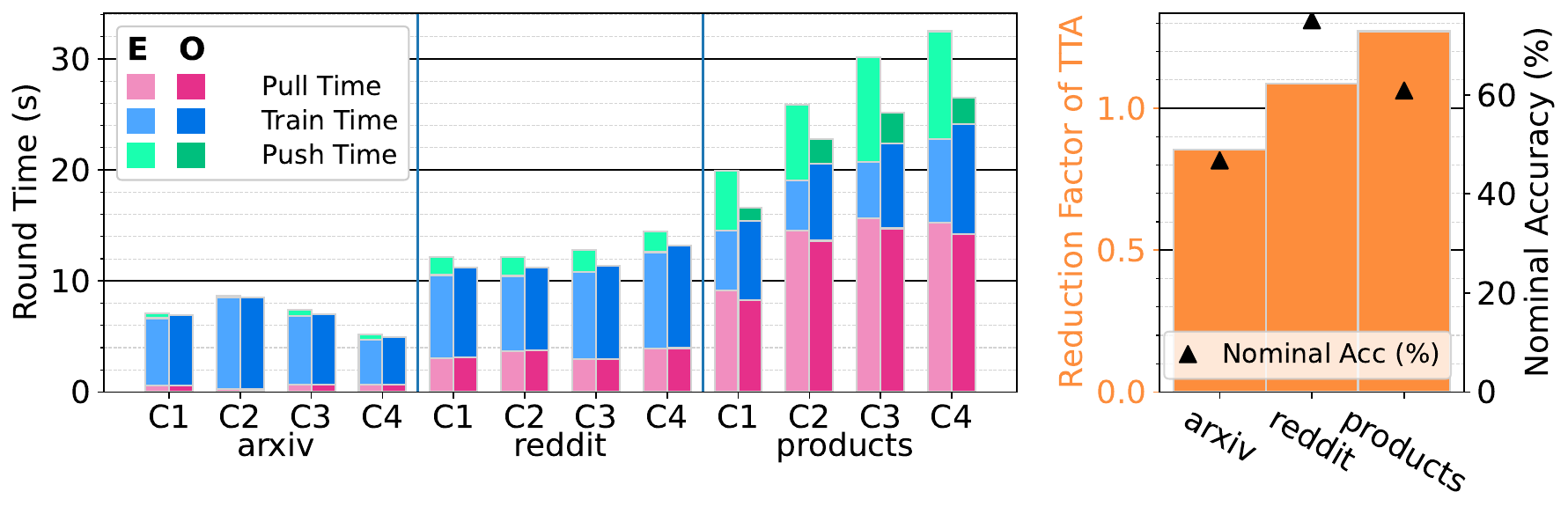}
    \caption{Effect of push phase overlap with training phase. Left plot shows the time taken in each phase for EmbC~(\textbf{E}) and OpES with Overlap~(\textbf{O}). Right plot shows the resulting reduction in time to accuracy (TTA).}
    \label{fig:tta-ovrl}
\end{figure}

Fig.~\ref{fig:tta-ovrl} illustrates the benefits of overlapping the pushing of embeddings with the final epoch training to hide the push latency. 
We contrast the time taken without and with the optimization in the left plot.
In general, we see that the push phase time~(green stack) in each client becomes negligible for Arxiv and Reddit and fairly small for Products when the overlap optimization is present~(right bar) compared to without the optimization~(left bar). 
But we also see a modest increase in the training time~(blue stack) since an additional concurrent push process partly competes with the training phase. Additionally, the push overlap optimization cannot fully mask the \textit{push time} in the case of Products~(green stack, right bar), as it does with Arxiv and Reddit. This is because the number of embeddings that ought to be shared by each client for Products is almost an order of magnitude larger when compared to the other two graphs~(see markers in Fig.~\ref{subfig-2:inside-outside}), and the training time is not sufficient to mask the time taken to push all of them. Even so, we see a substantial drop of $\approx3.8\times$ in push time averaged across the $4$ clients.

Due to the edge density~(Reddit) or~graph size (Products), these graphs end up with a large number of embeddings and, therefore, show the most reduction in the per-round times. 
This is also evident from the reduction in time-to-accuracy~(TTA) plot~(right plot in Fig.~\ref{fig:tta-ovrl}). 
Here, we plot the ratio of time taken without and with the optimization to achieve the same nominal accuracy, which is within 1\% of the smaller of the peak accuracies of these two approaches.
Values of $>1$ indicate that the overlap gives overall benefits. 
This not only encompasses the benefits of reduction in per-round time but also the reduction in accuracy due to sending stale embeddings from the $\epsilon-1$ epoch to bootstrap the next round of training.
The TTA is reduced by a factor of approximately $~1.1\times$ and $~1.3\times$ for Reddit and Products, respectively. Here, the large benefits of reduction in push time offset the marginal drop in accuracy per round.
However, in the case of Arxiv, the push phase itself takes little time, and its time reduction per round is small. In this case, the impact of stale embeddings is outstripped by the reduction in per-round time, and leads to an overall penalty in training time of 15\%~(TTA ratio of 0.85); overlap takes 5 rounds to reach this accuracy while non-overlap takes only 6 rounds, with the time per round being comparable~(left plot).

In summary, overlapping works well if the train time is a large fraction in a round and can hide the push time effectively, e.g., computationally intensive mini-batch training or a larger number of epochs per round, and where the downsides of stale embeddings~(which causes more rounds to converge) can be offset by faster rounds.

\begin{figure}[t!]
\centering
\begin{minipage}{0.47\textwidth}
\centering
\vspace{-0.65cm}
    \subfloat[arxiv\label{subfig:arxiv_pru}]{%
      \includegraphics[width=1\textwidth]{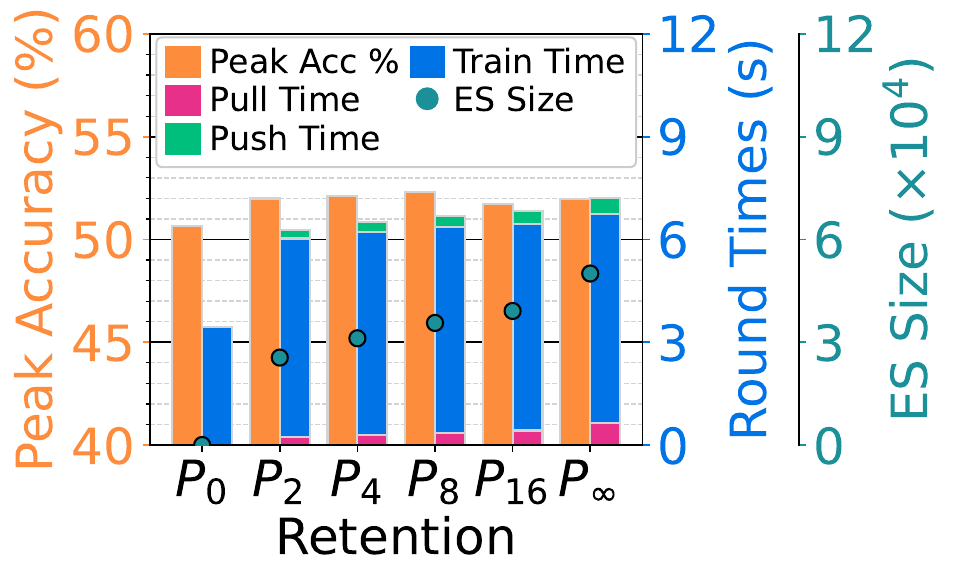}
    }\hfill
    \subfloat[reddit\label{subfig-2:reddit_pru}]{%
      \includegraphics[width=1\textwidth]{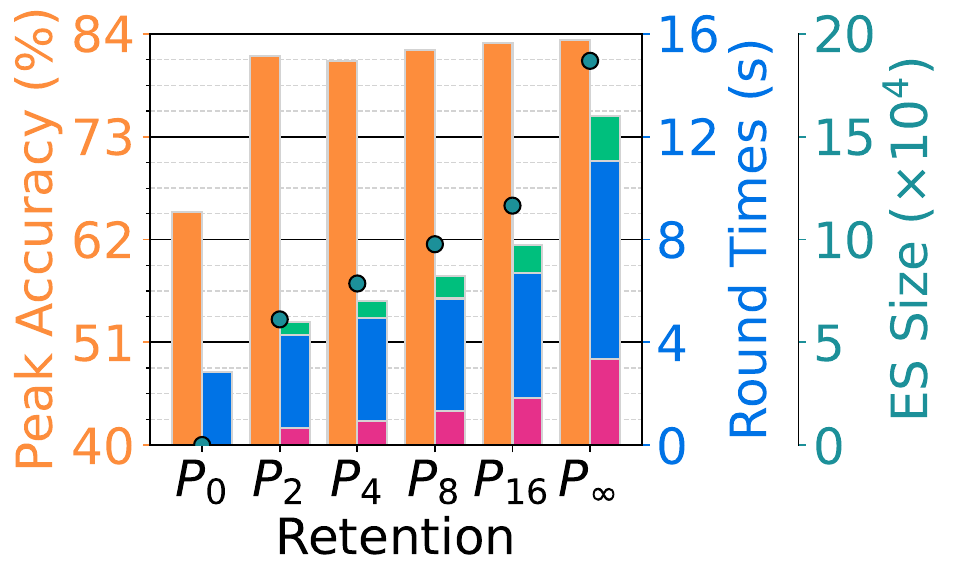}
    }\hfill
    \subfloat[products\label{subfig-3:products_pru}]{%
      \includegraphics[width=1\textwidth]{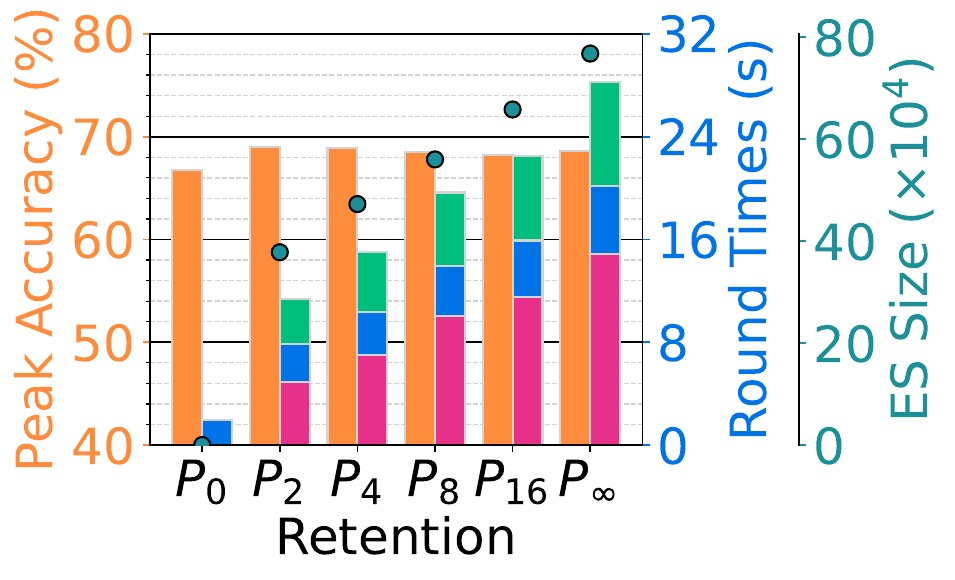}
    }
    \caption{Effect of different numbers of retained nodes during pruning~($P_i$) on per-round time, peak accuracy achieved and size of embeddings maintained for different graphs.}
    \label{fig:peak-acc-prune}
\end{minipage}\qquad
\begin{minipage}{0.42\textwidth}
\centering
    \subfloat[arxiv\label{subfig:arxiv}]{%
      \includegraphics[width=1\textwidth]{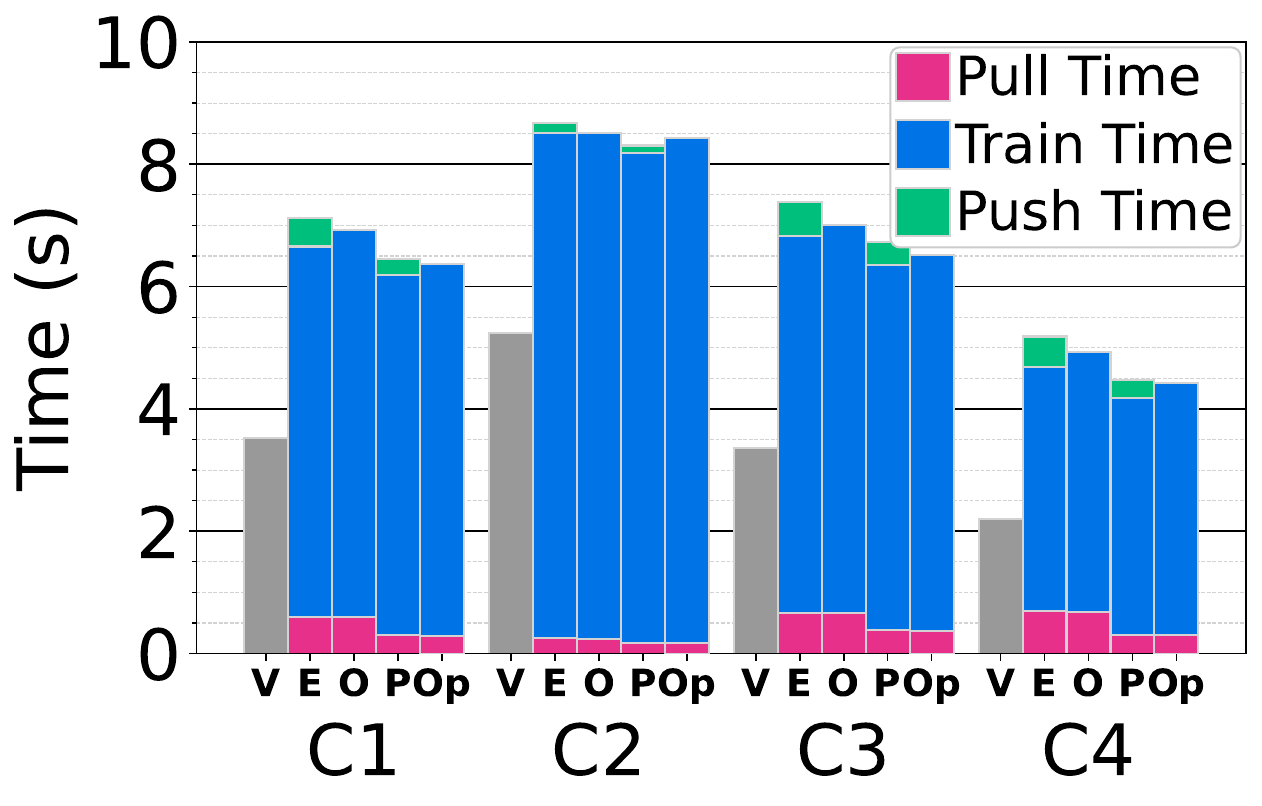}
    }\vspace{0.15cm}
    \subfloat[reddit\label{subfig-2:reddit}]{%
      \includegraphics[width=1\textwidth]{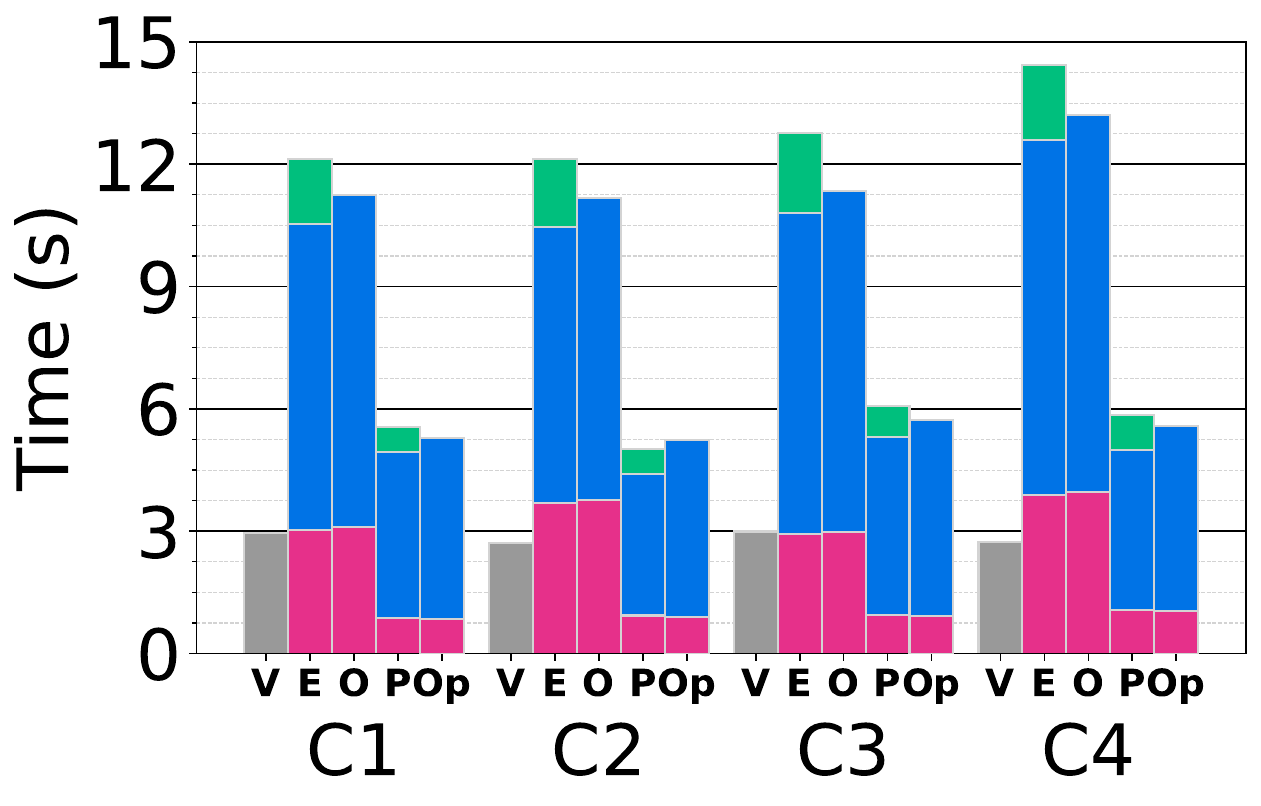}
    }\vspace{0.25cm}
    \subfloat[products\label{subfig-3:products}]{%
      \includegraphics[width=1\textwidth]{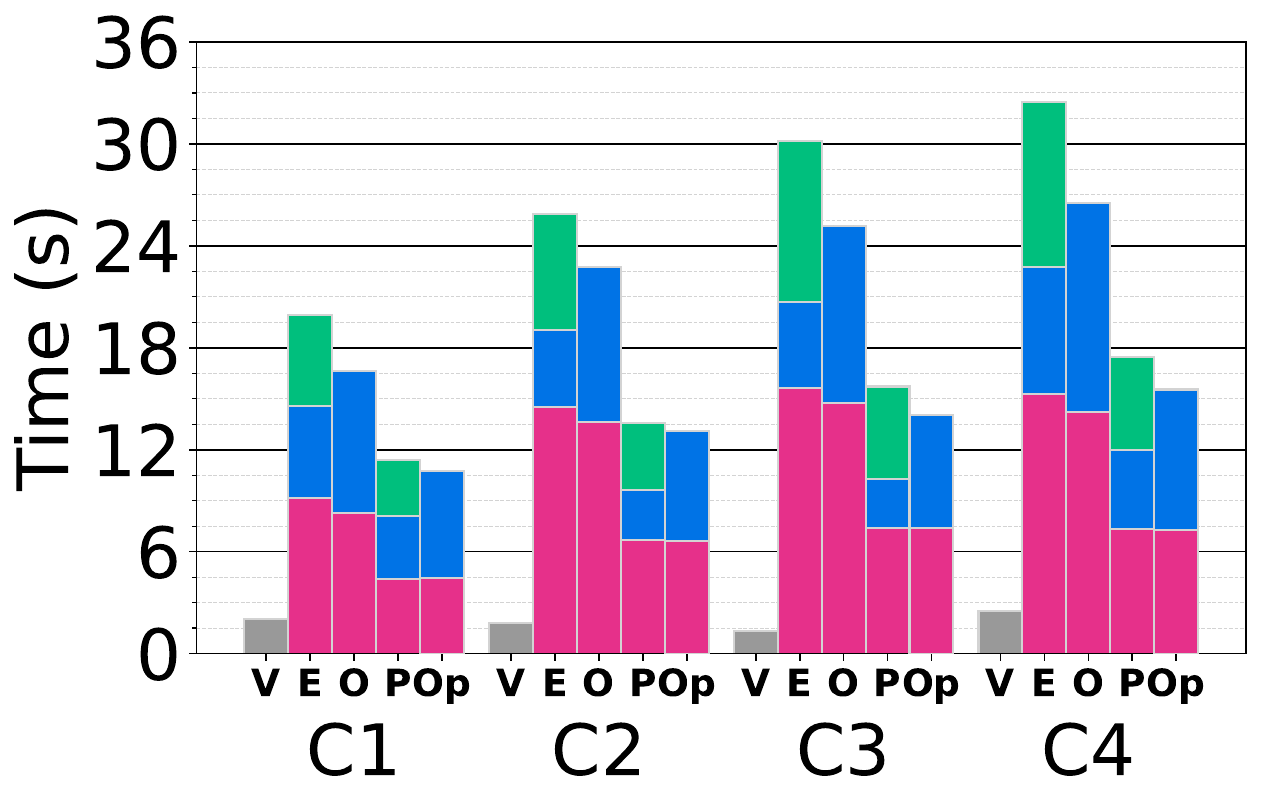}
    }
    \caption{Median per-round times for Vanilla federated GNN (\textbf{V}) and EmbC (\textbf{E}) baselines, and for \opes with Overlap (\textbf{O}), $P_4$ pruning (\textbf{P}) and Overlap+$P_4$ (\textbf{Op}) optimizations, for different graphs and clients.}
    \label{fig:client_wise_acc}
\end{minipage}
\vspace{-0.6cm}
\end{figure}
\subsection{Benefits of Pruning on Performance Optimization}

Fig.~\ref{fig:peak-acc-prune} shows the effect of pruning the expanded subgraph, using different \textit{retention limits} (maximum remote vertices retained, $P_i$), on the peak accuracy reached~(orange bar, left Y axis) and median per-round time~(stacked bar, right inner Y axis) showing time per phase as well.
We also report the number of embeddings maintained at the embedding server~(marker, right outer Y axis), which correlates with the push and pull times as well as the training time. $P_0$ indicates vanilla federated GNN where all remote vertices are pruned, while $P_\infty$ means no pruning optimization is done and all remote vertices are retained~(EmbC).

Pruning limits the expanded subgraph size, reducing the number of remote vertices for which embeddings are pulled/pushed in a round between the clients and the embedding server. 
As a result, we see significant reductions in the per-round time for Products and Reddit; by $2.7\times$ and $2.8\times$, respectively using $P_2$ compared to the default $P_\infty$. This is because their push times and pull times~(green and pink stacks, right inner Y axis) are higher;
however, the corresponding reduction in per-round time is modest ($1.2\times$) for Arxiv.

Both Reddit and Products have a much higher average degree compared to Arxiv~(Table~\ref{tab:dataset-specs}), causing them to have more remote vertices and, hence, more embeddings to pull/push. The general trend points to a sustained drop in embedding counts as the number of retained nodes reduces. Pruning each neighbourhood to the utmost 2 remote vertices~($P_2$) decreases the embedding counts by $3\times$ for Reddit and $2\times$ for Arxiv and Products. Reddit has a larger drop since it is much denser. 
Additionally, we notice a small decrease in the training time for each round in cases where fewer nodes are retained~(blue stacks, right inner Y axis). This behaviour is most likely due to the reduced number of nodes at each hop during the forward pass, for which the embeddings need to be loaded from the locally cached embeddings. This leads to a reduction in an expensive embedding matrix update operation during a forward pass.

\subsection{Comparative Performance with Baselines}

\begin{figure}[t]
\centering
    \subfloat[arxiv\label{subfig:arxiv-acc}]{%
      \includegraphics[width=0.33\textwidth]{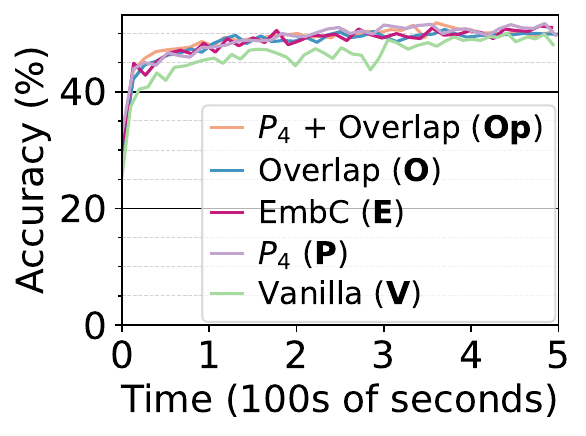}
    }
    \subfloat[reddit\label{subfig-2:reddit-acc}]{%
      \includegraphics[width=0.33\textwidth]{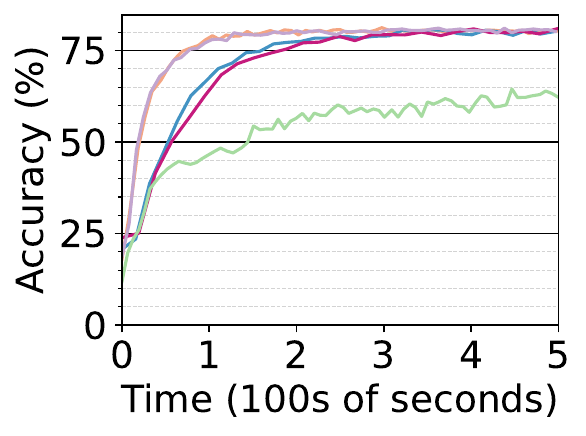}
    }
    \subfloat[products\label{subfig-3:products-acc}]{%
      \includegraphics[width=0.33\textwidth]{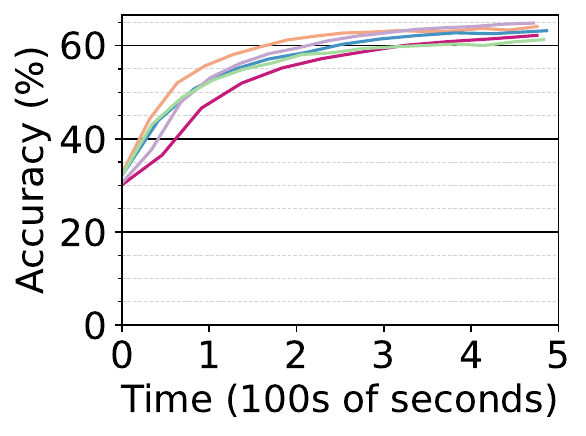}
    }
    \caption{Training timeline for all strategies~(clipped at 500s) for Vanilla federated GNN (\textbf{V}) and EmbC baselines~(\textbf{E}), and for \opes with Overlap (\textbf{O}), $P_4$ pruning (\textbf{P}) and Overlap+$P_4$ (\textbf{Op}) optimizations}
    \label{fig:acc-v-time}
\end{figure}

We compare our \opes optimizations with vanilla federated GNN~(VFL) and the state-of-the-art EmbC baseline. We report the per-round median times in Fig.~\ref{fig:client_wise_acc} and the training timeline till nominal convergence in Fig.~\ref{fig:acc-v-time}. The accuracies reported in the latter are based on server-side test datasets.
The retention limit is set to $P_4$. 

As expected, VFL has the smallest per-round time since it has no exchange of embeddings and only uses its local subgraph for training. However, it performs poorly for Reddit, which has a high edge degree and remote vertices, and losing access to remote embeddings reduces the accuracy to $62\%$, compared to EmbC and \opes, which achieve $\approx 80\%$ accuracy. However, for Arxiv and Products, VFL only has a modestly lower accuracy by a few \%. However, this is assuming an ``ideal'' partitioning using METIS that reduces the number of edge cuts and, hence, remote vertices. A more natural semantic partitioning expected in practical applications, e.g., based on vertex labels or ownership, can cause even Products or Arxiv to have more edge cuts and, consequently, worse performance for VFL.

For the SOTA baseline, EmbC without additional optimizations~(\textbf{E}), and our \opes with incremental optimizations~(\textbf{O, P, Op}), we observe the median round times on each client gradually reduce~(Fig.~\ref{fig:client_wise_acc}). As expected, the pull, push and train times decrease from EmbC~(\textbf{E}) to $P_4$~(\textbf{P}). Similarly, overlap~(\textbf{O}) masks the push times. Together, our \opes optimizations~(\textbf{Op}) reduce the per-round times by factors of $\approx 1.12\times$, $\approx 2.18\times$ and $\approx 2.2\times$ compared to EmbC, for the Arxiv, Reddit and Products graphs.

Fig.~\ref{fig:acc-v-time} are the corresponding convergence plots for the experiments reported in Fig.~\ref{fig:client_wise_acc}. Since Reddit is the densest graph of the three, it sees the most improvements using our optimizations and outperforms VFL and EmBC baselines, converging the fastest. This is because not only does it benefit the most from pruning on account of its density but also from the fact that its computationally intensive training round is able to mask the embedding push time effectively.
We observe that \opes also converges the fastest for Products because it, too, sees a sharp drop in per-round times with minimal detrimental effects of stale embeddings or pruning. Arxiv, on the other hand, does not show significant benefits in convergence for \opes over EmbC. As discussed before, Arxiv has low push times, which does not let the clients utilize the overlap benefit to the fullest. It is also sparse, which means not only is the information lost due to remote vertices not substantial, but it also does not benefit much from the pruning optimizations.

\section{Conclusion}
\label{sec:conclude}
In this paper, we presented Optimized Embedding Server (OpES), an optimized approach to the SOTA embedding-based federated GNN learning over subgraphs, like EmbC. \opes optimizations reduce the significant communications required in each federated learning round. We propose two optimizations, namely, overlapping of embedding communication with training and pruning of subgraph expansion.  This helps the embedding server scale with increased GNN depth and embedding size. We successfully show improvements in time-to-accuracy metrics and do an in-depth analysis of the parameters affecting the same. As future work, we plan to extend this framework to include support for dynamic and weighted pruning methods, pull time optimizations, and other improvements, such as clubbing of cross-client edges to reduce training costs further.

\bibliographystyle{plain}
\bibliography{arxiv-refs}
\end{document}